\documentclass[epj]{svjour}
\usepackage{graphics}
\begin{document}
\title{Modulated replica symmetry breaking schemes for
antiferrimagnetic spin glasses}
\author{R. Oppermann \inst{1}, D. Sherrington \inst{2}, M. Kiselev\inst{1}}
\institute{Institute for Theoretical Physics, University
of W\"urzburg, D-97074 W\"urzburg  \and Department of Physics, University of Oxford, Keble
Road, OX3NP UK}
\date{\today}
%
\abstract{
We define modulated replica symmetry breaking ($RSB$) schemes
which combine tree- and wave-like structures. A modulated scheme
and unmodulated $RSB$ are evaluated at 1-step level for a
semiconductor model with antiferromagnetic Korenblit-Shender interaction.
By comparison of the free energies we find evidence that a
$T=0$ phase transition in the ferrimagnetic phase
leads to a transition between the different RSB-schemes.
An embedding factor of Parisi block matrices with
sublattice-asymmetrical size is employed as a
new variational parameter in the modulated scheme.
} 
\titlerunning{Modulated replica symmetry breaking schemes}
\maketitle
\section{Introduction}
\label{intro}
Hierarchical tree-structures and replica symmetry breaking ($RSB$)
\cite{parisi1,parisi2} are celebrated features in the theory of
magnetic systems with random and frustrated interactions of infinite range.
Their role for short-range spin glasses was hotly debated recently
\cite{palassini,marinari}. A puzzling question over many years
concerned the existence of variants or alternatives for the Parisi
$RSB$-scheme. The latter proved to be very robust however; it is hard
to find relevant variables which perturb or change the scheme.

In this article we define and apply $RSB$-schemes which show a
wave-like modulation in addition to the tree-struc\-ture
\cite{parisi1,parisi2}. We work in the context of a two sublattice
infinite-range interaction model, defined and analyzed in
replica-symmetric ($RS$)-approximation by Korenblit and Shender
\cite{korenblitshender} ($KS$-model).
The $KS$-model successfully offered the description of transitions between
spin glass and antiferromagnetic order (or ferrimagnetic order in a field) in
spite of an infinite-range interaction. Spatially staggered order
is allowed by letting the interaction work only between different
sublattices. The field of application includes two-component magnets as well as
standard antiferromagnets, where staggered magnetic order defines sublattices.
Range-free interactions cannot distinguish spatial positions and consequently
unite the mean field picture of sublattice systems with another
class of systems having just an equal number of mutually interacting but
arbitrarily placed A- and B-spins.
The $KS$-model effectively mirrors the phase diagram of
the $SK$-model \cite{shk} to the antiferromagnetic side, still allowing to
retrieve ferromagnetic solutions. Intra-lattice interactions are a
less relevant detail \cite{korenblitshender}; they can yet be
included and dealt with in a refined $KS$-model. Transitions from
spin glass to ferrimagnetic order, driven by antiferromagnetic
interactions, are frequent physical phenomena and concern a wide
range of different microscopic models. Even in cases when quenched
disorder is weak or absent, spin glass models can have the power
to mimic behaviour of clean but geometrically frustrated systems
\cite{mezard,marinari2,chandra}. Beyond the present application to
antiferromagnetic instabilities and ferrimagnetic glassy phases,
our results suggest that the modulated $RSB$-schemes could also serve as a basis to
describe excited states in finite-range spin glasses.\\ \indent
\section{The two-sublattice spin glass model with competing antiferromagnetic- and
ferrimagnetic order}
\label{sec:1}
A class of Hamiltonians, for which the modulated RSB-scheme is constructed,
is given by the $KS$-model in an external field $h(r)$
\begin{eqnarray}
{\cal{H}}&=& -\sum_{i_A=1}^N\sum_{j_B=1}^N
J_{i_A,j_B}S(r_{i_A})S(r_{j_B})\nonumber\\
&+&\sum_{i_A} h(r_{i_A})S(r_{i_A})
+\sum_{i_B}h(r_{i_B})S(r_{i_B}),
\label{model}
\end{eqnarray}
where the partially frustrated random interaction $J_{iA,jB}$ is chosen to obey
a Gaussian distribution
$P(J_{iA,jB})=exp(-N(J_{iA,jB}+J_{af}/N)^2/(2 J^2))\sqrt{N/(2\pi J)}$.
Centered at a negative mean coupling $\langle J_{i_A,j_B}\rangle=-J_{af}<0$,
the sub\-lattice-interaction permits glassy antiferromagnetic order.
Glassy ferrimagnetic order with lifted $A\leftrightarrow B$ symmetry results
for example when a homogeneous field is applied or when spins of different lengths
(different spin quantum numbers in quantum models)
happen to be located on different sublattices.
Modulated $RSB$ should also be considered for model classes
including interactions of different types of localized spins,
for example tight-binding electron spins coupled to ionic spins $S$;
initially mobile carriers, which localize due to their interaction with
ionic spins, may not be able to fit the $RSB$ glassy order.
This example reaches far beyond the classical model (1).
In the present work we focus exclusively on model (1) with minimal inequivalence
of sites (such as being of $A$- and $B$-type), which requires a hybrid modulated
form of $RSB$ in a solvable classical model and hence reveals a
coupling of replica- and real space.

In replica theory \cite{virasoro}, which we use here, all spin variables
acquire a replica-index $a$, $S\rightarrow S^a$.
After elimination of the microscopic spins the corresponding effective Lagrangian
of the replica theory \cite{shk} is given in terms of Hubbard-Stratonovich fields
\cite{korenblitshender}.
The $SK$-model interaction requires one such field \cite{shk},
$\tilde{Q}^{a,b}$. Its statistical average
$Q^{ab}\equiv\langle \tilde{Q}^{ab}\rangle = \langle S^a_i S^b_i\rangle$
describes glassy order \cite{parisi1,parisi2}
in addition to a homogeneous magnetization $M=\langle S^a_i\rangle$, which can
be finite in case of partial frustration.
The $KS$-model however involves for each sublattice $\kappa=A,B$
a magnetization $M_{\kappa}\equiv\langle S_{i_{\kappa}}^a\rangle$ and
$Q^{ab}_{\kappa}\equiv\langle\tilde{Q}_{\kappa}^{a,b}\rangle$,
and a field $\tilde{Q}_3^{a,b}$ which couples the sublattices
\cite{korenblitshender}.
The averaged matrix $Q_{AB}\equiv -i\hspace{.1cm}Q_3\equiv -i\langle\tilde{Q}_3\rangle$
turns out to be equal to $Q_A+Q_B$, where $Q_A$ and $Q_B$ inevitably
show sublattice-splitting of their entries $q_A\neq q_B$ in {\it ferrimagnetic} phases,
together with $|M_A|\neq |M_B|$.
The {\it size} of their block-diagonal matrices, characterized in $RSB$ by a
Parisi parameter $m$ \cite{parisi1}, may also develop a sublattice-asymmetry.
Thus, at $1RSB$-level two order parameters for each sublattice $A$ or $B$, hence
$\{q_{1A}, q_{2A}, q_{1B}, q_{2B}\}$, denoting matrix elements of $Q_A$ and $Q_B$,
and two Parisi-parameters $m_A, m_B$ need to be considered.

\begin{figure}
\hspace{-.5cm}
\resizebox{.53\textwidth}{!}{%
  \includegraphics{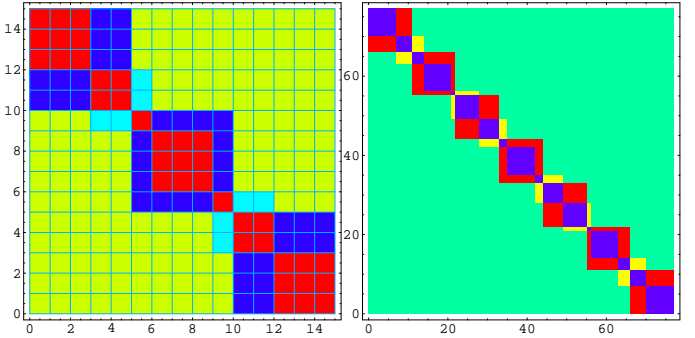}
}
\caption{Simple examples of modulated saddle point matrices $Q_{AB}=Q_A+Q_B$
with
$m_A=3$, $m_B=5$, and $n=15$ (left Figure) and $m_A=7$, $m_B=11$, and $n=77$
(right Figure) in one-step $RSB$. Four different regions are shown with entries
$q_{1A}+q_{1B}$, generated by the overlap-areas of $m_A\times m_A$- and
$m_B\times m_B$-sized block matrices along the diagonal,
with $q_{1A}+q_{2B}$ and $q_{2A}+q_{1B}$ generated by their nonoverlap-regions,
and $q_{2A}+q_{2B}$ belonging to the superposition of off-diagonal $A-$ and
$B$-elements.}
\label{modulatedq3}       
\end{figure}

A simple illustration for the matrix $Q_{AB}=-i\hspace{.1cm}Q_3$ is displayed in
Fig.\ref{modulatedq3}.
Note that for simplicity it is not shown that elements on the diagonal vanish
(while, for example in fermionic spin glasses, these elements equal $1$ at
half-filling; this detail can trivially be accounted for in the trace formulas
below, but is of no relevance for our present application).

%
%
{\it Unmodulated replica-symmetry breaking scheme}:
In the unmodulated 1RSB scheme one chooses $m_A=m_B$.
\\
\section{Modulated Replica Symmetry Breaking Schemes}
\label{MOD-RSB}
We consider the superposition of two diagonal Parisi-block matrices with
sizes $m_A\times m_A$ and $m_B\times m_B$ such that a rational embedding
factor $\gamma$ denotes the number of smaller blocks (let $m_A<m_B$)
{\it fully embedded} inside larger $m_B$-boxes, i.e. without being
intersected by the latter ones, divided (normalized) by the number of
larger boxes inside the entire ($n\times n$) host matrix
(the example of Fig.\ref{modulatedq3} show $\gamma=1$ (left figure)
and $\gamma=\frac57$ (right figure)).

We distinguish now a single embedding scheme, for which the host size $n$
is restricted to be the least common multiple of $m_A$ and $m_B$, and
multiple embedding schemes, which align $k$-times such structures
along the diagonal.

For all of $n, m_A, m_B, n/m_A,$ and $n/m_B$ integral and $m_A<m_B$,
the embedding parameter $\gamma$ is given by
\begin{equation}
\gamma=(n/m_A-(n/m_B-1))/(n/m_B).
\label{gamma}
\end{equation}
Let us now turn to the important replica limit $n\rightarrow0$.
\subsection{Single-embedding modulated scheme $($SMS$)$}
When the limit $n\rightarrow0$ is approached, as required in replica theory,
Eq.(\ref{gamma}) and the integral constraints associated with it must be
relaxed. This allows to obtain a finite nontrivial free energy.
To this end we choose $\gamma$ as a free variational parameter like $m_A$
and $m_B$. To obtain the free energy $F$ for this modulated $RSB$-scheme
we need the trace of the square of such super-imposed Parisi matrices
(see Eq.\ref{eq:application} below).
The free energy involves the limit as $n$ goes to zero of (1/n) times
a sum of traces over the $Q^2$-values. As for standard Parisi matrices
\cite{parisi1,parisi2}, $tr\{Q_{\kappa}^2\}=n(m_{\kappa}-1)q_{1\kappa}^2
+n(n-m_{\kappa})q_{2\kappa}^2, \hspace{.1cm} \kappa=A,B$,
so for a physical finite free energy one requires that $tr\{Q_3\}^2$ also
scales like $n$ for small $n$.
A set of self-consistent equations is derived by the condition
that $m_A, m_B$ {\it and} $\gamma$ extremize the free energy in
the replica limit.
The idea is thus to find a function of $m_A, m_B$ and $\gamma$,
which agrees with the trace of $Q_{AB}^2$ for integer-valued parameters
$m_A, m_B, n$ allowed by the scheme, satisfies the integral constraints
ahead of Eq.(\ref{gamma}),
and whose analytical continuation for small $n$ is linear in $n$.
This goal is achieved by evaluating all overlap contributions to $tr\{Q_3^2\}$
and reexpressing the results for each of the four different overlap regions
in terms of the embedding factor $\gamma$.

The trace formula for $Q_3^2$ for arbitrary block-sizes $m_A < m_B$,
which all nest the host-matrix of size $n$,
for the single-embedding scheme, is expressible in terms of $\gamma$ as
\begin{eqnarray}
%
& &\Xi(\gamma)\equiv tr\{Q_{AB}^2\}=n(m_A-1)(q_{1A}+q_{1B})^2\nonumber\\
& &+n(m_B-m_A)(q_{2A}+q_{1B})^2+n(n-m_B)(q_{2A}+q_{2B})^2\nonumber\\
& &+2(q_{1A}-q_{2A})(q_{1B}-q_{2B})\phi(\gamma),
\label{eq:3}
\end{eqnarray}
where the dependence on the embedding factor $\gamma$ is contained in
\begin{eqnarray}
& &\hspace{-.8cm}\phi(\gamma)\equiv\frac{n}{3m_B^3}\{(m_B-n)(m_B-(1+\gamma)m_A)
\nonumber\\
& &\hspace{-.5cm}[(m_B-2n)m_B+m_A(2n(1+\gamma)-(4+\gamma)m_B))]\}.
\label{eq:4}
\end{eqnarray}
\vspace{-.2cm}

The result represented by Eq.(\ref{eq:3}), together with $\phi$
given by Eq.(\ref{eq:4}), holds for all integral and non-integral (rational)
values of $\gamma$ allowed by the construction.
Let us consider a few examples, using this division into two classes
with either integral or non-integral embedding factors $\gamma$:

i) there exists a subset of matrices, where each $m_B$-block hosts the
same (integral) number of $m_A$-blocks. The left hand side of
Fig.\ref{modulatedq3} shows one example with
$(m_A=3, m_B=5, n=15, \gamma=1)$. Further examples of this class
are $(2,7,14,3)$, $(4,7,28,1)$, $(3,8,24,2)$;

ii) the right hand side of Fig.\ref{modulatedq3} presents one example for
the matrix-class having non-integral embedding factors with
$(m_A=7, m_B=11, n=77, \gamma=5/7)$. Let us add further examples by
$(3,4,12,2/3)$ and $(7,17,119,11/7)$.

One can see that $\Xi$-contributions from all overlap regions can be
expressed in terms of $m_A$, $m_B$, $\gamma$, while the number of these overlap
regions depends explicitly on the host matrix size $n$. This feature
guarantees the finite replica limit of the free energy.

\subsection{Multiple-embedding modulated scheme $($MMS$)$}
\begin{figure}
\begin{center}
\resizebox{.27\textwidth}{!}{%
  \includegraphics{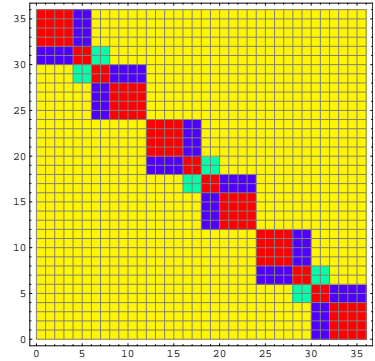}
}
\end{center}
\caption{Example of a modulated saddle point matrix belonging to
the multiple-embedding scheme:
$\langle Q_3\rangle/i=\langle Q_A\rangle+\langle Q_B\rangle$ with
$m_A=4$, $m_B=6$, and $n=36$
in one-step $RSB$.}
\label{mms-example}
\end{figure}

We also define a modulated scheme $MMS$ which incorporates a $k$-fold repeated
$SMS$-structure (of size $n_1\times n_1$) along the diagonal of an
$n\times n$ host matrix, for example $(4,6,24,3/4)$ where $k=2$.
The $SMS$-matrix size is chosen as a variational parameter,
kept finite while the replica limit $n\rightarrow 0$ is taken,
and finally varied to extremize $F$.
Altogether $m_A, m_B$, and $n_1$ are variational parameters, which
determine the embedding factor $\gamma(m_A,m_B,n\rightarrow n_1)$
according to Eq.(\ref{gamma}), or $m_A, m_B$ and $\gamma$ are
varied and their selfconsistent solutions yield
$n_1(\gamma)=m_A m_B/(m_A-m_B+m_A\gamma).$
The free energy is obtained by means of $\lim_{n\rightarrow0}
\Xi(n,n_1)/n$, where
\begin{equation}
\Xi(n,n_1)=\Xi(\gamma(n_1))n/n_1+
n(n-n_1)(q_{2A}+q_{2B})^2
\label{Eq.:4b}
\end{equation}
or $\Xi(n,\gamma)=\Xi(\gamma)n/n_1(\gamma)+ n(n-n_1(\gamma))
(q_{2A}+q_{2B})^2$ in case $\gamma$ is varied.
An example for a three-fold embedding is given by Fig.\ref{mms-example}.
Since relation (\ref{gamma}) is part of the
definition of the $MMS$ and since the replica limit maps the Parisi-type
parameters $m_A,m_B, n_1$ from $[1,\infty]$ into the interval $[0,1]$,
$\gamma$ is thus restricted to values $\gamma>-1$ in contrast to the $SMS$.
The $MMS$ also differs essentially from the $SMS$ by the fact that the
(A,B)-symmetric limit $m_A=m_B\equiv m$ reduces it to the unmodulated $1RSB$-scheme.
Another possible variant of $MMS$, where the number $k=n/n_1$ of repeated
$SMS$-structures is varied, is discarded, since $k$ cannot extremize $F$.
\subsection{Upgrade of the modulated RSB-schemes}
While the modulated $RSB$-schemes are initially designed for applications to
glassy ferrimagnetic systems, one should also consider them under a more
general point of view:
they can be used in systems without sub-lattices and even
without reference to antiferro- or ferrimagnetic order. One can
upgrade the schemes as an alternative of the Parisi-scheme
provided the traces of higher powers $Q^k, k\geq 3$, also yield a
non-divergent replica-limit.
This separate point as well as higher order $RSB$ is not the issue of the
present article, where only $tr Q_3^2$ is needed for the $KS$-model.
%
%
\section{Application to a layer model with a confined magnetic field}
\label{application}
In order to examine the specific features of the modulated $SMS$-scheme
in comparison with the unmodulated one, we performed a detailed analysis for
the $KS$-model (1) in a spatially confined magnetic field $H_p$.
For an equal number of $\alpha N$ of $A$- and $B$-spins we chose
$h=H_p$ and $h=0$ for the remaining $(1-\alpha)N$ spins.\\
The free energy of this model realization in 1-step $RSB$ can be decomposed
into three parts
\begin{equation}
\label{eq:application}
F=F_0+\alpha F_1(H_p)+(1-\alpha)F_1(H_p=0),
\end{equation}
where
\begin{equation}
\hspace{-.2cm}F_0=
-J_{af}M_A M_B-
\frac{J^2}{4T}\left[ lim_{_{_{\hspace{-.5cm}n\rightarrow0}}}\frac1n
tr \underline{Q}^2
-2\sum_{\kappa}(1-q_{1\kappa})\right],
\end{equation}
and
\begin{equation}
\hspace{-.2cm}F_1(H_p)=-\sum_{\kappa=A,B}\frac{T}{m_{\kappa}}\int^G_{z_{2\kappa}}
\ln\int^G_{z_{1\kappa}}
\cosh^{m_{\kappa}}\{\frac1T\tilde{H}_{\kappa}(H_p)\}
\end{equation}
with $\int_z^G\equiv \int_{-\infty}^{\infty}dz\hspace{.05cm}
e^{-z^2/2}/\sqrt{2\pi}$, and $\underline{Q}\equiv(Q_A,Q_B,Q_3)$.\\
The effective field $\tilde{H}_{\kappa}$ on sublattice $\kappa$
depends on the order parameters of the complementary sublattice $\bar{\kappa}$.
It is given in terms of 1. the confined polaron field $H_{\kappa}$,
2. in terms of the magnetization field of the complementary sublattice
$\bar{\kappa}$, and 3. spin fields representing the spin glass field
$z_{k,\kappa}$ (in the $1RSB$-discretized approximation), by the expression
\begin{equation}
\tilde{H}_{\kappa}(H_p)=
H_p-J_{af}M_{\bar{\kappa}}+J\sqrt{q_{2\bar{\kappa}}}\hspace{.1cm}
z_{2\kappa}+ J\sqrt{q_{1\bar{\kappa}}-q_{2\bar{\kappa}}}
\hspace{.1cm}z_{1\kappa},\nonumber
\label{eq:heff}
\end{equation}
The motivation for choosing this specification of the $KS$-model is
essentially twofold:

1) by scanning the full range $0<\alpha<1$ we found that the phase
diagram is not only marked by a continuous spin glass -
ferrimagnet transition. At slightly higher ratios $J_{af}/J$, a subsequent
small flop transition from ferrimagnet to an (what might be called)
antiferrimagnetic phase occurs,
which provides an ideal test-ground for the $SMS$-scheme:

the $A\leftrightarrow B$-symmetries are strongly broken, since solutions
are far away from either $M_A=M_B$ or $M_A=-M_B$. The
main features of the phase diagram for all $\alpha$ at selected characteristic
polaron fields $H_p$ and temperatures is analyzed below in \ref{sec:phdiag}.

2) $CdTe/Cd_{1-x}Mn_x Te$-layers are well described by the present
model, where $H_p$ represents a confined polaron field created by
polarized exciton-hole spins being localized at the interface.
The penetration depth of the hole-wavefunction defines
the portion $\alpha$ of the $CdMnTe$-layer which is exposed to the
field $H_p$.
The magnetic $CdMnTe$-layer employs Villain-Ising pseudo-spins $S$
\cite{villain}
representing tetrahedra of manganese Heisenberg-spins which retain only two
orientational degrees of freedom. All pseudospins of the magnetic
layer are then coupled by a long-range partially frustrated interaction with
antiferromagnetic mean value. This model (with $\alpha=0.5$) provided
optimal fits for experiments in the
spin glass regime at $x=33\%$ \cite{chudnik}.
Increasing Mn-concentration $x$ enhances the antiferromagnetic bias and
eventually leads to a transition from spin glass to antiferromagnetic or
ferrimagnetic order (in a homogeneous polaron- or external field)
at a critical concentration $x_c$.
Scaling and numerical analysis on the basis of anisotropic Heisenberg models
were also provided \cite{rigaux,zippelius}. The virtue of the
Pseudo-Ising concept lies in the smaller lower critical dimension when
compared to Heisenberg systems.

We explore at $T=0$ the difference between the $SMS$-scheme and the unmodulated one.
For its demonstration we choose a polaron-field strength $H_p=4 J$ and a
confine\-ment-fraction $\alpha=0.5$ (half-penetrated layer).
%
%
\begin{figure}
\resizebox{.5\textwidth}{!}{\includegraphics{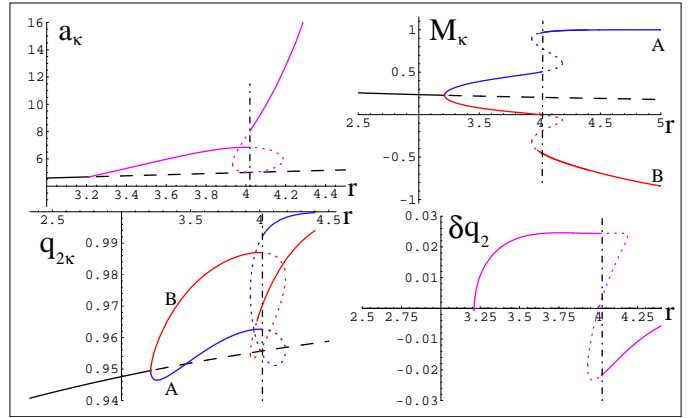}}
\caption{Solutions for $T=0$ and $\alpha=0.5$ in 1-step unmodulated $RSB$:
Parisi parameter $a_A=a_B$, $a_{\kappa}=m_{\kappa}/T$, sublattice
magnetizations $M_A,M_B$, spin glass order parameters $q_{2A},
q_{2B}$, and $\delta q_2\equiv q_{2B}-q_{2A}$, as a function of
$r\equiv J_{af}/J$ and finite confined field $H_p=4 J$.
Dash-dotted lines locate the 1st-order transition, dashed lines
show unstable solutions.}
\label{oldscheme}
\end{figure}
%
%
\begin{figure}
\resizebox{.5\textwidth}{!}{\includegraphics{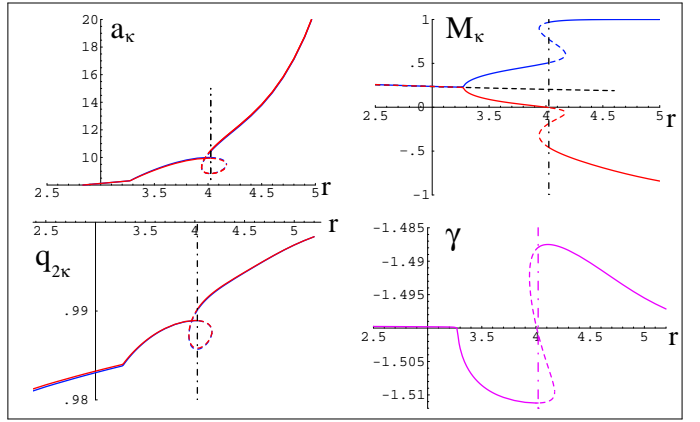}}
\caption{Corresponding results in the modulated
$RSB$-scheme $SMS$ showing ($a_A,a_B$), ($M_A,M_B$), ($q_{2A},q_{2B}$), and
the $SMS$-embedding parameter $\gamma$. It is seen that $a_{\kappa}$ and
$q_{\kappa}$ show only very weak sublattice splitting.}
\label{newscheme}
\end{figure}
%
%
Our 1-step $RSB$-results are obtained by solving up to seven coupled
selfconsistent integral equations ($SMS$-scheme) which extremize $F$.
Thanks to the $T\rightarrow0$-limit one integration can be solved exactly,
simplifying the selfconsistent set of ten coupled double-integral
equations for finite $T$ considerably
(while $q_{1A}=q_{1B}=1$ at $T=0$, both $q_{1\kappa}(T)$, $\kappa=A,B$, and the
hole polarization too must be determined selfconsistently for finite $T$).
The $T=0$-results of Figs.\ref{oldscheme}, \ref{newscheme} show
a continuous SG-ferrimagnetic transition with order parameter $M_A-M_B$
to occur at $r\equiv J_{af}/J\approx 3.25$,
followed by a discontinuous transition to antiferrimagnetic order near
$r\approx 4.02$. As Figure \ref{oldscheme} shows, $q_{2A}$ and $q_{2B}$ undergo
large jumps and become (almost) interchanged at the discontinuous transition in
the framework of the standard unmodulated scheme, ie under the condition
$a_A=a_B$, $a_{\kappa}\equiv lim_{_{T\rightarrow 0}} m_{\kappa}/T$.

In the modulated SMS-scheme, Fig.\ref{newscheme}, the
selfconsistent solutions for $a_A, a_B$ and $q_{2A}, q_{2B}$ are different
but show only small sublattice splitting; at the transition $a_B-a_A$ changes sign,
in contrast to $q_B-q_A$.

The sublattice effective fields of Eq.\ref{eq:application} help
to explain the origin of the discontinuous transition: the competition
between $J_{af}M_{\bar{\kappa}}$ and the polaron field $H_p$ leads
(in the clean limit) eventually to a total spin reversal on one sublattice.
Random magnetic order would smear the jump in any homogeneous field
but the discontinuity reappears due to the competition between
antiferromagnetic (\{$A,B$\}- symmetric) order, preferred in the $H_p=0$-region,
and a strongly ($A,B$)-asymmetric ferrimagnetic order for sufficiently large
$H_p/J$.

We finally compare our $1RSB$-results for the free energies in the double transition
regime of Figs.\ref{oldscheme},\ref{newscheme}.
\begin{figure}
\resizebox{.5\textwidth}{!}{%
  \includegraphics{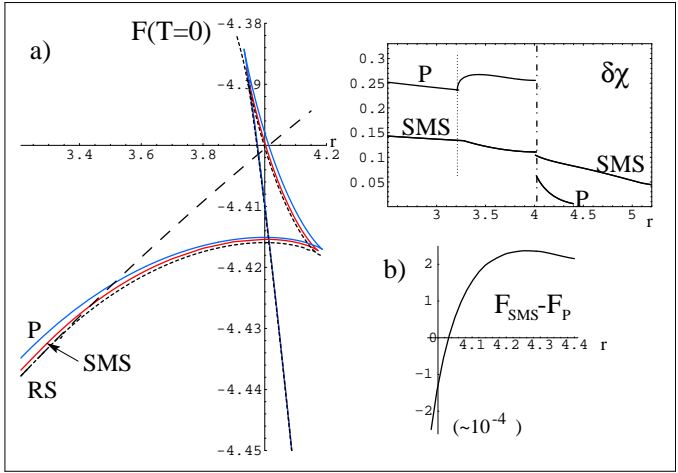}
}
\caption{a) Energies $F(T=0)$ of modulated $1RSB$- (arrow),
unmodulated $1RSB$- ($P$), and $RS$-scheme ($RS$) with
ferromagnetic $RS$-solution (dashed), shown as a function of
$r\equiv J_{af}/J$ in the double transition region, b) energy
difference between modulated- and unmodulated scheme; inset (top
right) shows susceptibility $RSB$-contributions $\delta\chi$.}
\label{freeenergy}
\end{figure}
Fig.\ref{freeenergy} provides evidence for the discontinuous transition
to involve a transition from unmodulated to modulated $RSB$.
As discussed in \cite{parisi1} higher energies correspond to improved solutions
(unless identical stability criteria are met).
Fig.\ref{freeenergy}b) shows that the energy for the $SMS$-scheme is higher for
$J_{af}>J_{af}^c\approx 4.02$ (lower if $<$ holds).
This crossing of energies at the discontinuous transition and
the character of the $RSB$-schemes suggests that modulated $RSB$ governs
the $J_{af}>J_{af}^c$-regime. The $RS$-solution is lowest but
unstable everywhere.
Despite small energy separation, equilibrium- and nonequilibrium linear
susceptibility shows large differences which depend strongly on
the type of $RSB$-scheme (inset of Fig.\ref{freeenergy}).
Further observable manifestations and thermal behavior remain
to be considered. \\




\section{Confined field-fraction effect on the multiplicity of phase transitions}
\label{sec:phdiag}
\begin{figure}
\resizebox{.5\textwidth}{!}{%
  \includegraphics{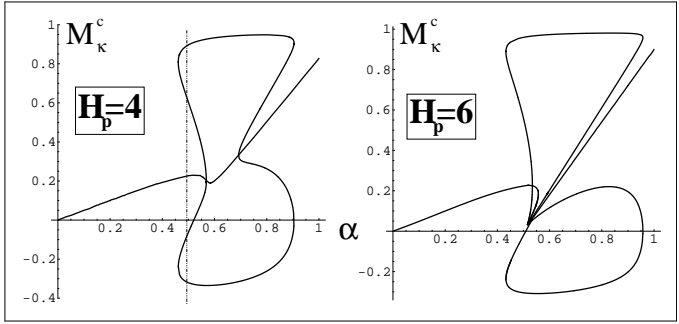}
}
\caption{Magnetizations $M^c_{\kappa=A,B}$
obeying the constraint
$dM^c_{\kappa}/dr=\infty$, $r\equiv J_{af}/J$,
are shown in $RS$-approximation for $T=0$ as a function of the polaron fraction
$\alpha$ for field strengths $H_p=4,6$; the vertical line
indicates the position of the double transition analyzed in
$1RSB$-schemes in Figs.\ref{oldscheme},\ref{newscheme}.}
\label{phdiag}
\end{figure}

The sublattice effective fields of Eq.\ref{eq:application} help
to explain the origin of the discontinuous transition: the competition
between $J_{af}M_{\bar{\kappa}}$ and the polaron field leads
(in the clean limit)
to a total spin reversal on one sublattice. Random magnetic order would
smear the jump in any homogeneous field, as can be deduced
from Fig.\ref{phdiag} at $\alpha=1$,
but the discontinuity reappears due to the competition between
antiferromagnetic (($A,B$)-symmetric) order, preferred in the $H_p=0$-region,
and a strongly ($A,B$)-asymmetric ferrimagnetic order for sufficiently large
$H_p\neq 0$.
For half-penetrated layer ($\alpha=0.5$) the continuous
transition with order parameter $M_A-M_B$,
caused by the competition between spin glass and antiferromagnetic order,
exists at $J_{af}=J_{af}^c\approx 3.25 J$.
To understand the $\alpha=0.5$-scenario in the context of all
$0\leq\alpha\leq 1$ we found sufficient to analyze the stability
limits at $T=0$ in an $RS$-approximation.
Introducing the definitions
\begin{equation}
\xi=r \hspace{.1cm}M_{\kappa}, h_p\equiv H_p/J
\end{equation}
one may cast the equation of state into the compact nested form
%
\begin{equation}
\xi=u(u(\xi)+\eta\hspace{.1cm} u(\xi-h_p))+\eta\hspace{.1cm} u(u(\xi)+
\eta\hspace{.1cm} u(\xi-h_p)+h_p),
\end{equation}
%
where
%
\begin{equation}
u(\xi)\equiv (1-\alpha)r\hspace{.1cm}erf(\xi/\sqrt{2}), \eta =\alpha/(1-\alpha).
\end{equation}
The stability limits are obtained under the constraint
\begin{equation}
\frac{dr(M_{\kappa})}{dM_{\kappa}}=0,
\frac{d^2r(M_{\kappa})}{dM_{\kappa}^2}\neq 0.
\end{equation}
%
in terms of the inverted solution $r(M_{\kappa})$ displaying the interaction ratio
$r\equiv J_{af}/J$ as a function of the sublattice magnetizations $M_{\kappa}$.
The constrained solutions $M^c_{\kappa}$ are derived by scanning all
$\alpha$ and shown in Fig.\ref{phdiag} for typical values of $H_p$
(stability limits in terms of $J_{af}(\alpha)$ are omitted for brevity).
At a fixed $\alpha$, the existence of one or three solutions implies
a single continuous
or a single discontinuous SG-ferrimagnetic transition respectively,
while five solutions are necessary to obtain a double transition regime
(with a magnetization-curve $M(r=J_{af}/J)$ shaped
as in Figs.\ref{oldscheme},\ref{newscheme}).
Corrections in the effective field $\tilde{H}$ originating from
intra-sublattice interactions do not change qualitatively the results.
%
%
The right part of Fig.\ref{phdiag} shows a special point which emerges
for large $H_p$ near $\alpha\approx 0.5$ and small $M^c_{\kappa}$.

The Korenblit-Shender model in a field $h=H_p$ can be retrieved at $\alpha=1$.

\section{Conclusions and open ends}
In this article hybrid $RSB$-schemes which combine tree- and
wave-like structures were defined.
A variational embedding factor of Parisi block matrices appeared as
a characteristic ingredient of the new schemes.
The application showed that the $SMS$-modulated phase is preferred
beyond a critical $J^c_{af}$ where a type of antiferrimagnetic order prevails.
The $MMS$-scheme (in contrast to $SMS$) allows a continuous crossover to
unmodulated $RSB$ and should be analyzed as a candidate for glassy
antiferromagnets.
A generalization of our trace-formula to all powers of the order
parameter matrix can create an extension of the Parisi scheme,
modelling perhaps excited states in short-range spin glasses.
Higher orders of both types of $RSB$ are currently under study,
using for example techniques applied to unmodulated $RSB$ in
Ref.\cite{os-prb}.

\section{Acknowledgements}
We acknowledge support by the EPSRC (R.O. and D.S.), by the DFG, the
SFB410 {\it II-VI semiconductors} (R.O. and M.K.),
and by the ESF-programme SPHINX. We express our gratitude (R.O. and D.S.)
for hospitality at MPI Heidelberg and for discussions with
H.A. Weidenm\"uller.

%
%

\end{document}